# Metastable giant moments in Gd-implanted GaN, Si, and sapphire

X. Wang[1*], C. Timm[2], X. M. Wang[3], W. K. Chu[3], J. Y. Lin[4], H. X. Jiang[4], and J. Z. Wu[5]

[1]Department of Pharmaceutical Chemistry, University of Kansas, USA

[*]Email: xiangw@ku.edu

[2]Institute of Theoretical Physics, Technische Universtät Dresden, 01062 Dresden, Germany

[3]Department of Physics, University of Houston, USA

[4]Department of Electrical and Computer Engineering, Texas Tech University, USA

[5]Department of Physics and Astronomy, University of Kansas, USA

## Abstract

We report on Gd ion implantation and magnetic characterization of GaN films on sapphire substrates and of bare sapphire and Si substrates to shed light on the mechanism underlying the induced magnetism upon Gd ion implantation. For all three hosts, giant magnetic moments per Gd ion were observed at temperatures of 5 through 300 K. The maximum moment per Gd in GaN was 1800 $\mu_B$, while the moments in Gd-implanted Si and sapphire were only slightly smaller. The apparent induced ferromagnetic response was found to be metastable, disappearing after on the order of 50 days at room temperature, except for the implanted sapphire. We argue that our findings support a defect-based picture of magnetism in Gd-implanted semiconductors and insulators.

# 1. INTRODUCTION

In recent years, diluted magnetic semiconductors (DMS) [1,2] have been studied extensively, partially driven by the hope for applications such as the integration of data processing and non-volatile data storage on a single microelectronic chip. DMS are produced by introducing magnetic ions into originally non-magnetic semiconductors. Ferromagnetic response has been observed for many DMS. A long-range magnetic interaction between the local moments mediated by conduction-band or valence-band carriers is usually assumed to be responsible. Unfortunately, the Curie temperature $T_C$ in reasonably well understood compounds such as GaAs doped with Mn is limited to below 200 K [3,4].

On the other hand, many groups have reported ferromagnetism above room temperature in other DMS, mostly in wide-gap nitride and oxide semiconductors. The magnetism in these compounds is found to be fickle: samples grown following essentially identical protocols may show very different magnetic response including the presence or absence of apparent ferromagnetic order [2,5].

At first glance, Gd-doped GaN films grown by molecular beam epitaxy (MBE) and Gd-ion-implanted GaN films belong to this class since they are wide-gap DMS showing ferromagnetic response at high temperatures [6,7,8,9,10,11]. However, this system is even more puzzling: Dhar *et al.* [7,8,10] and Khaderbad *et al.* [11] have shown that at very low Gd concentrations down to 7 x $10^{15}$/cm$^3$, magnetic hysteresis is still seen at room temperature and the magnetic moments per Gd ion are enormous, up to 4000 $\mu_B$. This is observed both for MBE-grown and for ion-implanted GaN samples. These giant magnetic moments, which correspond to 4000 unpaired electron spins per Gd ion, are not well understood. On the other hand, Hejtmánek *et al.* [12] have observed ferromagnetic response of Gd-implanted GaN at temperatures up to 700

K, but did not find giant moments.

Dhar *et al.* [7,8] propose that the GaN host material is spin-polarized in the vicinity of the Gd ions. A phenomenological model based on this idea fits the data of the saturation magnetization vs. Gd concentration reasonably well. However, the spin polarization of a perfect GaN host crystal would be negligible: The exchange interaction between Gd *f*-electrons and host electrons is small compared to the GaN band gap of 3.47 eV [13,14], which is approximately the energy required to generate two unpaired spins. Dhar *et al.* [8] note that the large ionic radius of $Gd^{3+}$ should lead to strong strain, which, together with the large piezoelectric coefficient of GaN, might generate an electric potential that could trap electrons close to Gd. While this may lead to a few unpaired electron spins in the vicinity of a Gd ion, it is not a plausible source of thousands of spins.

*Ab-initio* calculations by Dalpian and Wei [15] suggest that localized empty minority-spin levels derived from Gd *f*-states are present in the gap. If the host material contains additional shallow donors with donor levels above the Gd-induced states, electrons would drop into these states. If the concentration of these donors is high enough, a large magnetization opposite to the Gd *f*-shells may result. The overlap between these polarization clouds could lead to a ferromagnetic interaction between Gd moments as in the bound-magnetic-polaron (BMP) picture [16]. However, it is again difficult to understand how this mechanism could produce thousands of localized states per Gd ion.

Nevertheless, the additional dopants invoked in [15] are likely important. GaN typically contains a high concentration of defects, in particular nitrogen vacancies [17,18], which act as shallow donors [19,20] and may be responsible for the n-type background carriers in GaN. GaN films on sapphire also contain a high concentration of dislocations [21]. A neutral single donor is expected to carry an unpaired spin 1/2. The total saturation magnetization of the MBE-grown GaN:Gd samples of Dhar *et al.* [7,8] appears to be the sum of a Gd-concentration-independent contribution and a contribution proportional to the Gd concentration, except perhaps for the sample with the lowest concentration. It is natural to attribute the constant contribution to defects. An estimate from the observed magnetization [7,8] gives a defect concentration of 1.8 x $10^{20}$/cm$^3$ for wurtzite GaN, which is at the upper limit for not intentionally doped GaN [18].

Magnetic moments contributed by defects have been implicated for the magnetism in various metal oxides such as ZnO and SnO$_2$ doped with magnetic ions as well as in oxides without magnetic doping [22,5]. Long-range magnetic order could then result from a BMP picture [23] or from a Stoner instability in the defect-induced impurity band [24,25]. Supercell calculations within density functional theory by Liu *et al.* [14] predict a strong ferromagnetic interaction between Gd *f*-moments and the moments of holes introduced by intrinsic gallium vacancies, whereas the interaction with electrons introduced by nitrogen vacancies is weak.

The idea of defect-induced magnetism is supported by experiments. For example, Kaspar *et al.* [26] find that the ferromagnetism in Cr$_x$Ti$_{1-x}$O$_2$ anatase correlates with the presence of structural defects. Yoon *et al.* [27] have been able to tune the ferromagnetism by intentionally introduced oxygen vacancies and structural defects in anatase TiO$_2$. Moreover, it has been demonstrated that ion implantation with non-magnetic ions can lead to ferromagnetic response in oxides [28,29,30] and Si [31]. Since ion implantation for GaN introduces defect levels into the gap [32], a defect-based mechanism is also plausible for Gd-implanted GaN. Following this idea, Khaderbad *et al.* [11] have studied the effect of rapid high-temperature annealing of Gd-implanted GaN. For the sample implanted at a low dose of 2.4 x $10^{11}$/cm$^2$ they observe a reduction of the saturation magnetization by roughly 50%, consistent with part of the defect-induced moments being annealed out. Further rapid annealing did not change the magnetization. The authors [11] suggest that the magnetic moments are carried by gallium and nitrogen interstitials that occur with high concentrations at the end of range of the implantation. On the other hand, the sample implanted at a high dose of 1.0 x $10^{15}$/cm$^2$ shows no change upon rapid annealing and its magnetization is consistent with being carried by Gd *f*-moments. Ney *et al*. [33] have found partially metastable ferromagnetic response and memory effects in MBE-grown GaN:Gd and InN:Cr and attribute these observations to defect-induced moments.

The lack of understanding of the ferromagnetic response in Gd-doped and Gd-implanted GaN suggests to study other samples from independent sources and compare them to other host materials in order to clarify the conditions for the ferromagnetic signals. We here report on magnetic characterization of GaN films grown on sapphire substrates and ion-implanted with relatively low concentrations of Gd. We compare the results to the bare sapphire substrate and to Si wafers, both implanted with Gd.

## 2. EXPERIMENTAL

The GaN epilayer was grown on the c-plane (0001) of sapphire substrates by metal organic chemical vapor deposition (MOCVD). The growth began with a thin GaN buffer layer, followed by the GaN epilayer grown at about 1050 °C. The film thickness is about 2 μm and electrical resistivity is in the range of 100 – 200 Ω cm. High crystalline quality of the material was confirmed with a narrow linewidth of 300 – 400 arcsec from the (002) rocking curve in x-ray diffraction. The lattice constants are *a* = 3.19 Å and *c* = 5.19 Å. Electrical transport measurements showed that the sample was n-type with a background electron concentration typically in the range of 7 x $10^{16}$/cm$^3$ and mobility of 650 cm$^2$/Vs [34].

Gd$^+$ implantation was performed using a 200 keV NEC ion implanter at The Texas Center for Superconductivity at the University of Houston. 190 keV Gd$^+$ ions were implanted into GaN at room temperature tilted 10° off-axis to avoid channeling. The dose ranged from $10^{11}$/cm$^2$ to $10^{14}$/cm$^2$. The current at 10 nA on the sample was scanned over a circular aperture of 15 mm diameter. A simulation of the Gd profile and defect distribution was performed using the commercial software SRIM. The result is summarized in Fig. 1. It should be noted that the implanted Gd$^+$ ions and the implantation-generated defects are both located in a thin surface layer of the sample with a thickness around 60 – 70 nm and with the peak locations shifted by only 10 – 12 nm in depth. Since all Gd ions are within a 100 nm deep surface layer, the average Gd ion density is estimated at around 1.2 x $10^{19}$/cm$^3$. Similar results have also been obtained from SRIM simulations for Gd ion-implanted sapphire and Si. The main results are summarized in Table 1, which shows the location of the peak concentration of Gd ions and defects generated by the irradiations. The close proximity of the implanted Gd$^+$ ions and the irradiation-generated defects may facilitate defect-induced magnetism.

The magnetic properties of the samples have been characterized using a Quantum Design SQUID Magnetometer. The samples were mounted in a plastic straw and secured by thread. The surface of the samples was always maintained perpendicular to the applied magnetic field *H*. The magnetization *M* as a function of *H* was measured at various temperatures *T*. In order to eliminate any spurious signal generated by a remnant field from the last measurement, the SQUID magnetometer was zeroed before each measurement. In addition, the background magnetization in applied field generated by the straw and the thread was subtracted from all the measurements.

## 3. RESULTS AND DISCUSSION

Figures 2(a) and 2(b) show the magnetization of a GaN film at various temperatures without and with Gd ion implantation, respectively. A background magnetization linear in applied field has been subtracted from the data. The dose of implantation was 1.2 x $10^{14}$/cm$^2$. For comparison, both sample areas are normalized to 1 cm$^2$. This normalization is also applied for the rest of the paper where similar comparisons are made. Figure 2 shows that a large magnetization was induced by the implantation. Subtracting the data of Fig. 2(a) from Fig. 2(b), we obtain the net magnetization contribution $\Delta M$ originating from the implantation; the result at $T = 5$ K is plotted in Fig. 3(a). The resulting average magnetic moment per Gd ion is about 1.67 x $10^{-17}$ emu at 5 K, which corresponds to about 1800 $\mu_B$. Here the magnetic moment per Gd ion is denoted by $\Delta M$ in units of $\mu_B$ and the measured $\Delta M$ as a function of temperature is plotted in Fig. 3(b). We thus confirm the observation of Dhar *et al*. [10] and Khaderbad *et al*. [11] of giant magnetic moments induced by Gd implantation at temperatures of 2 through 300 K. The results of Dhar *et al*. [10] for similar doses are shown as open symbols in Fig. 3(b). Based on an interpolation between the data of Dhar *et al*. [10] to our dose, we find that the induced giant magnetic moments at a given dose are roughly one order of magnitude larger in the present study. Besides on a GaN film, Gd-ion implantation was also conducted on a p-type Si wafer and on a bare sapphire substrate with the same dose. We have also observed giant induced magnetic moments in these cases. For Si, the induced $\Delta M$ per Gd ion is about 700 $\mu_B$ and for sapphire, about 500 $\mu_B$, both at 5 K.

Several of the samples were remeasured after being kept at room temperature in a dry box for various amounts of time. Fig. 4 shows magnetization curves of the Gd-implanted GaN and Si samples measured between 4 and 66 days after the implantation, as well as the magnetization curves of the non-implanted samples. From Fig. 4(a) we can see that the original induced magnetization in implanted GaN has almost disappeared after 66 days. This time-dependent effect is also observed for the implanted Si wafer, Fig. 4(b). Specifically, at a field of 5000 Oe, 46.7% of the induced magnetization had disappeared after 15 days and 88.8%, after 66 days. Interestingly, we find the magnetization to be stable for Gd-implanted sapphire over a time span of 132 days, as shown in Fig. 5.

To summarize our main findings, we have confirmed the observation of Dhar *et al.* [7,8,10] and Khaderbad *et al.* [11] of giant magnetic moments per Gd in GaN at temperatures of 5 through 300 K. Our results contradict the recent work by Hejtmánek *et al.* [12], who did not observe giant moments. While it is not fully clear which properties of the samples are responsible for the presence or absence of giant moments, we note that the doses used by Hejtmánek *et al.* [12] were higher than the ones used by Dhar *et al.* [10] and by us. Moreover, we also find apparent ferromagnetic signals with giant magnetic moments in Gd-implanted Si and sapphire. The moments are on the same order of magnitude as for GaN. The presence of a magnetic response for Si and sapphire suggests that the effect is not specific to a certain host. In particular, the size of the band gap and the strength of the spin-orbit coupling are clearly not crucial, since the three materials differ strongly in these respects. On the other hand, the apparent universality of the effect is consistent with the idea of defect-based magnetism, since ion implantation generates a high concentration of defects in any of the hosts.

The magnetization in applied field in implanted GaN was found to disappear over 66 days of room-temperature annealing, whereas the signal in implanted Si was strongly reduced over the same period, and implanted sapphire shows no clear annealing effect. The effect for GaN and Si is more pronounced than the 50% reduction of the magnetization by rapid high-temperature annealing in implanted GaN [11]. The decay of the signal shows that the entities carrying the magnetic moments are metastable and are slowly annealed out at room temperature. This strongly supports a defect-based mechanism. We conclude that the room-temperature annealing of the relevant defects is much slower in sapphire compared to GaN and Si. This is consistent with the general observation that in comparison to Si and compound semiconductors, the ion-irradiation-generated defects in oxides are the most difficult to anneal.

The present understanding of magnetism in semiconductors implanted with rare-earth ions leaves several open questions. The distribution of Gd and of implantation-induced and other defects is not known at the atomic scale. In addition, the nature of the defects mostly responsible for magnetism is not clear. Nitrogen and gallium vacancies as well as nitrogen and gallium interstitials are possible candidates. Formation of clusters may play a role [5]. The relevant annealing mechanism is not known but our experiments show that it is efficient at room temperature in both Gd-implanted GaN and Si.

In GaN, the exchange interaction of Gd $f$-moments with holes is predicted to be stronger than with electrons [14], based however on calculations for a single intrinsic defect per Gd instead of roughly a thousand, as relevant here. Hole-mediated magnetism is hard to reconcile with our experiment for implanted n-type GaN, since it would require the low-dose implantation to change the material to p-type. However, the likely weak interaction between Gd $f$-moments and electrons in states derived from the conduction band [14] should lead to a weak electron-mediated interaction between Gd moments inconsistent with ferromagnetic correlations at room temperature. Related to this, the nature of the long-range interaction is not understood. Is it BMP-like or do Stoner correlations in an impurity band play a role? The universality of giant magnetic moments in very different compounds implanted with Gd suggest to search for a universal mechanism.

## Acknowledgments

Fruitful discussions with S. Dhar and P. Saxena are gratefully acknowledged. This work was supported by NSF and by the University of Kansas General Research Fund Allocation No. 2301060. GaN growth work at TTU is partially supported by DOE (Grant No. DE-FG02-09ER46552).

**Figure captions**

Figure 1: Gd profile and defect distribution in the GaN film based on a simulation using SRIM 2008. Here, Gd$^+$ ions are accelerated to 190 keV and the implantation is carried out at room temperature. The displacement per atom (DPA) is defined as fluence multiplied by $V_{vac}/N_{at}$, where $N_{at}$ = 8.8467 x $10^{22}$ atom/cm$^3$ and $V_{vac}$ is the distribution from SRIM.

Figure 2: Magnetization $M$ as a function of applied magnetic field $H$ of a GaN film (a) without Gd implantation and (b) with Gd implantation at various temperatures from 5 to 300 K. The sample areas have been normalized to 1 cm$^2$ for the purpose of comparison.

Figure 3: Net magnetization $\Delta M$ induced by Gd implantation at a dose of 1.2 x $10^{14}$/cm$^2$ (a) as a function of applied magnetic field $H$ at 5 K and (b) as a function of temperature, in units of $\mu_B$. The results of Dhar *et al.* [10] for Gd doses of 1.1 x $10^{13}$/cm$^2$ and 1.0 x $10^{15}$/cm$^2$ are also plotted in (b) for comparison.

Figure 4: Magnetization $M$ at 5 K as a function of applied magnetic field $H$ of (a) the GaN film and (b) a Si wafer measured at various times after the Gd implantation. The sample areas have been normalized to 1 cm$^2$ for the purpose of comparison. The magnetization of a control sample without Gd implantation is also plotted for both cases.

Figure 5: Magnetization $M$ at 5 K as a function of applied magnetic field $H$ of the sapphire substrate measured 20 and 132 days after the implantation.

**Table caption**

Table 1: Gd and defect peak positions and peak concentrations for implanted GaN, sapphire, and Si. All results are simulations generated using SRIM 2008.

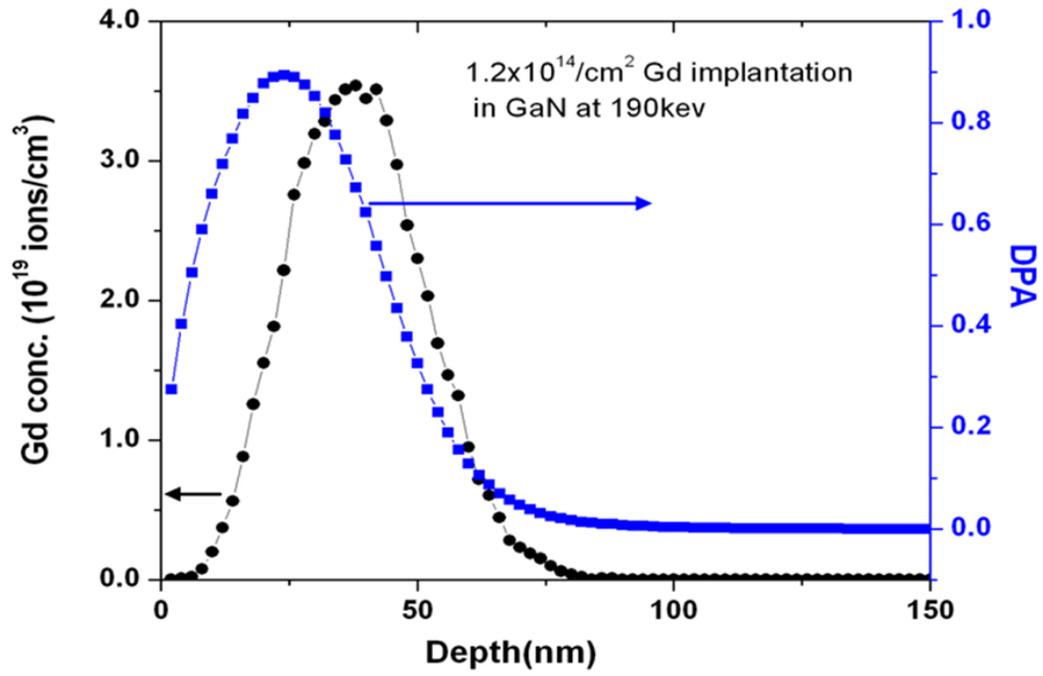

Fig. 1

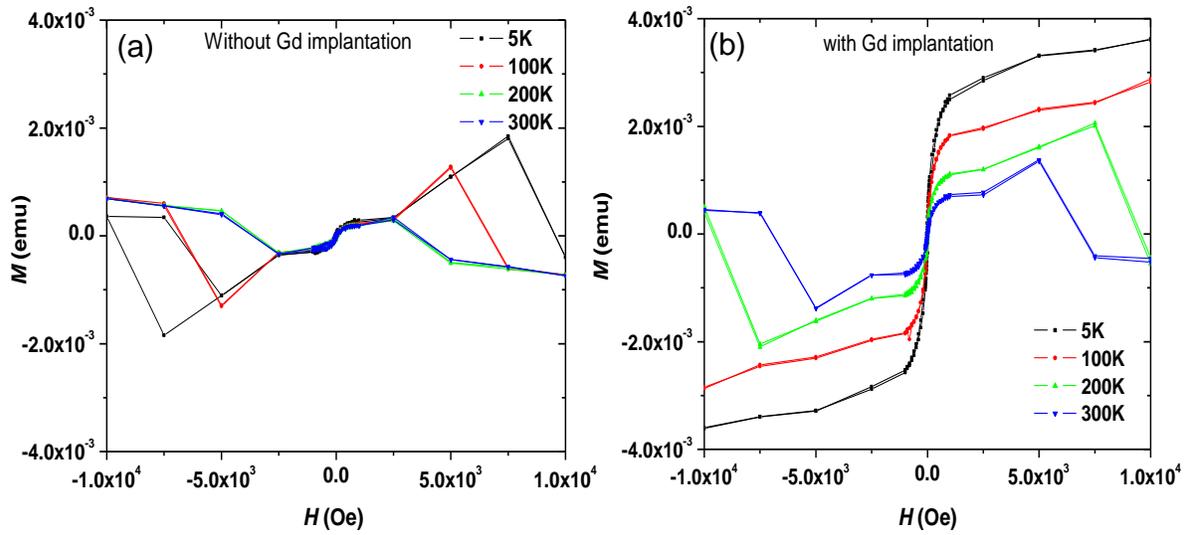

Fig. 2

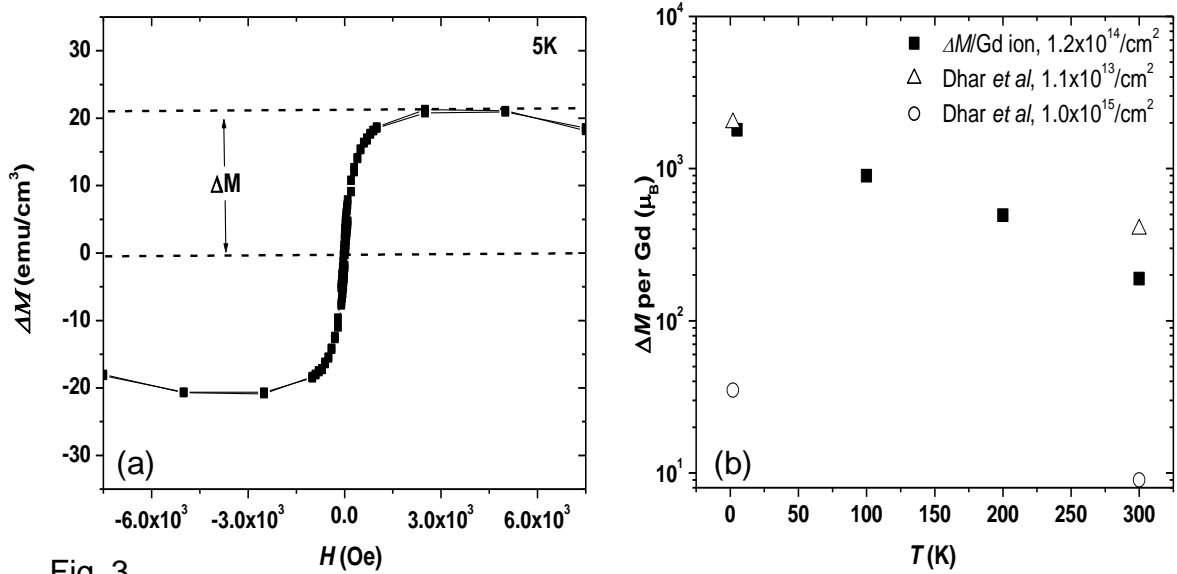

Fig. 3

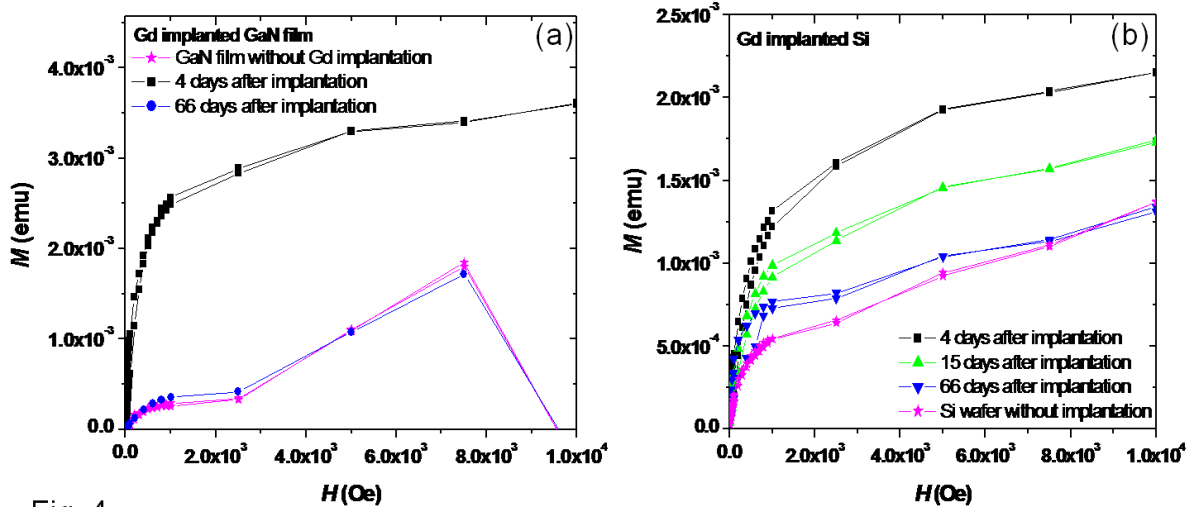

Fig. 4

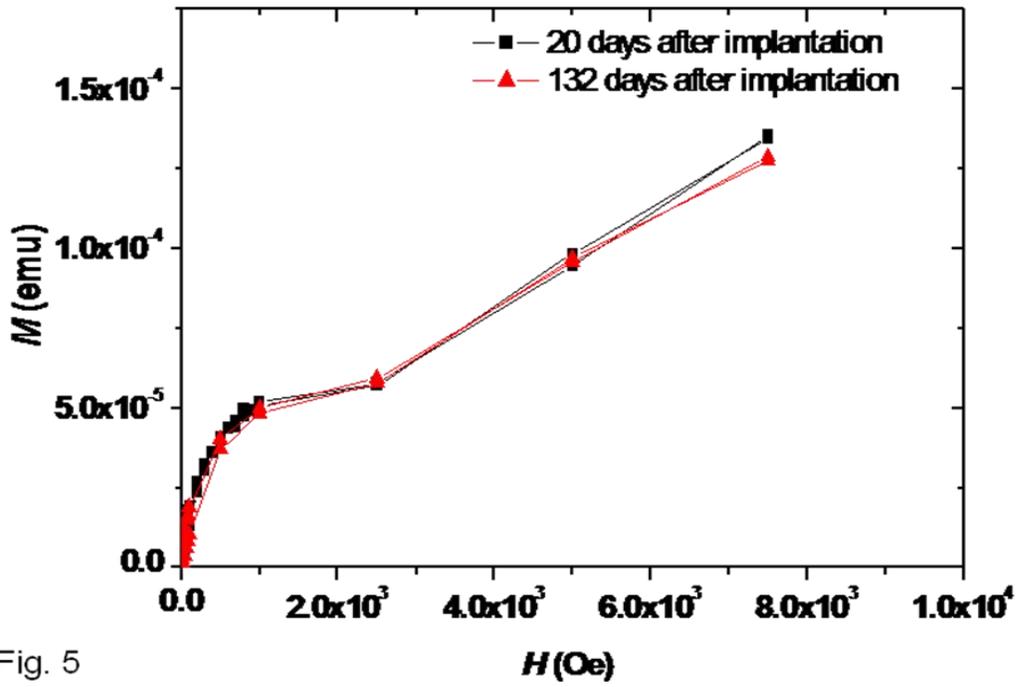

Fig. 5

|  | GaN | Al$_2$O$_3$ | Si |
|---|---|---|---|
| Gd peak position (nm) | 37.7 | 41 | 75 |
| Gd peak concentration (10$^{19}$ions/cm$^3$) | 3.6 | 5 | 2.4 |
| DPA peak position (nm) | 22 | 28 | 48.4 |
| DPA peak concentration | 0.93 | 0.65 | 1.392 |

Table 1